\documentclass[12pt]{article}
\usepackage{eurosym}
\usepackage[left=1in, right=1in, top=0.75in, bottom=0.75in]{geometry}
\usepackage{hyperref}
\usepackage{mathtools}
\usepackage{graphicx}
\usepackage{mathrsfs}
\newcommand{\bra}[1]{\bigl\langle #1 \bigr|}
\newcommand{\ket}[1]{\bigl| #1 \bigr\rangle}
\newcommand{\expect}[1]{\left\langle #1 \right\rangle}
\begin{document}

	\renewcommand{\baselinestretch}{1.3} \topmargin=-1.8cm \textheight=23 cm
	\textwidth=23cm
	
	\begin{center}
 {\large The  orthogonality speed of two-qubit state interacts locally with spin chain in the presence of
Dzyaloshinsky-Moriya interaction}\\
\vspace{0.7cm}
D. A. M. Abo-Kahla$^{a,b}$, M. Y. Abd-Rabbou$^{c}$, and N. Metwally$^{d,e}$\\
\vspace{0.3cm}
$^a$ Department of Mathematics, Faculty of Science, Taibah University, KSA.\\
$^{b}$ Math. Dep. , Faculty of Education, Ain Shams University, Cairo, Egypt.
\newline
$^{c}$Math. Dep., Faculty of Science, Al-Azhar University, Nasr City 11884, Cairo.\newline
	$^{d}$ Department of Mathematics, College of Science, University of Bahrain, Bahrain. \newline
	$^{e}$ Department of Mathematics, Aswan University, Aswan, Sahari 81528, Egypt.

\end{center}

\begin{abstract}
The orthogonality time is examined for different initial states settings interacting  locally with different types of spin interaction:
 $XX$, Ising and anisotropic models.
It is shown that, the number of orthogonality  increases, and consequently the time of  orthogonality decreases as the environment qubits increase.
The shortest time of orthogonality is displayed for the $XX$ chain model, while the largest time is shown for the Ising model. The external field increases
the numbers of orthogonality, while Dzyaloshinsky-Moriya interaction decreases the  time of orthogonality. The initial state settings together with the external field
has a significant effect on decreasing/increasing the time of orthogonality.

\end{abstract}

\section{Introduction.}
Quantum speed is a significant tool for developing the quantum computer. Its importance lies with the physical limits levied by the quantum mechanical laws on the speed of the quantum information processing and transfer of this information between users  \cite{margolus1998maximum,lloyd2000ultimate}. Quantum speed limit is based on the Heisenberg uncertainty principle associated with the time and energy, by which the minimum time that an initial state needs to evolve to its final state is determined. This determination is crucially important for quantum technology  \cite{PhysRevLett.103.240501}, such as, quantum communication \cite{PhysRevLett.46.623,PhysRevA.82.022318}, quantum computation  \cite{lloyd2000ultimate,PhysRevLett.110.050403}, and metrology \cite{giovannetti2011advances}. In other words, the minimum time prerequisite for a quantum system to pass from one orthogonal state (node) to another is called the speed of orthogonality  \cite{PhysRevA.72.032337}. It was investigated for a single two-level atom (single qubit) interacting with a rectangular pulse \cite{metwally2012information}, and interacting quantized field \cite{metwally2008controlling,obada2011quantum}.

On the other hand, the exchange interaction of the antisymmetric Dzyaloshinsky–Moria (DM) was introduced by Dzyaloshinsky \cite{dzyaloshinsky1958thermodynamic}.  Several studies have discussed the qubit system in
the Heisenberg spin chain. For example, the quantum speed limit of a quantum system consist  a central spin coupled to a Heisenberg spin-1/2 XY chain has been studied \cite{wei2016quantum}.  For a finite XY spin chain coupled to the external magnetic field the quantum speed limit has been discussed \cite{hou2017quantum}.  Also, quantum speed has been studied for a central qubit interacting with the isotropic Lipkin-Meshkov-Glick environment  \cite{hou2016quantum}. The effect of dynamical dissociate on the speed-up role by applying the strong "bang-bang" (BB) pulses control on the central spin system has been examined  \cite{wei2016dynamical}. addition, the criticality of quantum speed limit for a coupled system of a two-qubit system in the Heisenberg spin chain interacting with the antisymmetric Dzyaloshinsky-Moria (DM) has been explored  \cite{yin2019quantum}.

Our motivation in this work is to investigate the behavior  of the quantum speed orthogonality of a two qubit system  coupled to a Heisenberg XY spin chain in the presence of an external magnetic field and  DM interaction. The initial state is prepared in generalized Werner state. So, we have organized this paper as a follows: In Sec. 2, we present the analytical solution of the physical model,
and we obtain the final density operator. In Sec. 3, we discuss the quantum speed orthogonality in Werner state. In Sec. 4, we discuss the quantum speed orthogonality in a maximum entangled state. The final section is devoted to present our conclusion.

\section{Physical Model:}
Let us consider a two qubit system that is coupled to a Heisenberg XY chain in the presence of DM interaction in an external field. The total Hamiltonian of this system is $H_{T}=H_{E}+H_{I}$, with;
\begin{equation}
	H_{E}=-\sum_{l}^{N}\biggl\{\frac{1+\gamma }{2}\sigma _{l}^{x}\sigma
	_{l+1}^{x}+\frac{1-\gamma }{2}\sigma _{l}^{y}\sigma _{l+1}^{y}+\lambda
	\sigma _{l}^{z}+D(\sigma _{l}^{x}\sigma _{l+1}^{y}-\sigma _{l}^{y}\sigma
	_{l+1}^{x})\biggr\},
\end{equation}
and
\begin{equation}
H_{I}=\frac{-g}{2}(\sigma _{A}^{z}+\sigma
_{B}^{z})\sum_{l}^{N}\sigma _{l}^{z}.
\end{equation}
where, $H_{E}$ and $H_{I}$ are representing the Hamiltonian of the environment, and the interaction between the two qubit system and the spin–chain environment. $l$ is the site of the chain, $N$ is the total number of sites, and $\sigma _{l}^{i}, i=x,y,z$ are the Pauli matrices. The factor $\gamma$ ($ 0<\gamma<1$)is the anisotropy of exchange interaction in the $xy$ plane. $D$ ($ -1\leq D\leq1$) is factor of the DM interaction  along z direction, and $g$ describes the coupling between the system
and surrounding environment chain.
By using the Jordan--Wigner transformation, and replacing $\sigma _{j}^{x}=\sigma _{j}^{+}+\sigma _{j}^{-}$, $ \sigma _{j}^{y}=i\left( \sigma _{j}^{-}-\sigma _{j}^{+}\right)$, and $\sigma _{l}^{z} =1-2n_{l}$. The energy Hamiltonian $H_{E}^{\lambda _{\nu }}$ may be rewritten as \cite{Cheng2009}:

\begin{eqnarray}
H_{E}^{\lambda _{\nu }} &=&-\sum_{l}^{N}\left(
C_{l+1}^{+}C_{l}+C_{l}^{+}C_{l+1}\right) +\gamma \left(
C_{l}^{+}C_{l+1}^{+}+C_{l+1}C_{l}\right)
\nonumber\\
&&+2iD\left( C_{l+1}^{+}C_{l}-C_{l}^{+}C_{l+1}\right) -2\lambda _{\nu
}C_{l}^{+}C_{l}+\lambda _{\nu },
\end{eqnarray}
where $\sigma _{l}^{+}=\prod_{j=1}^{l-1}(1-2n_{j})C_{l} $, $\sigma _{l}^{-}= \prod_{j=1}^{l-1}(1-2n_{j})C_{l}^{+}$, $n_{j} =C_{j}^{+}C_{j}$, and $C_{l}^{+}$ ($C_{l}$) is fermionic creation (annihilation) operator. However, $\lambda _{\nu } $ are the eigenvalues ($\lambda _{1}=\lambda+g, \lambda _{2}=\lambda=\lambda _{3}, \lambda _{4}=\lambda-g,  $)

The energy environmental Hamiltonian $H_{E}^{\lambda _{\nu }}$ can be written under the Fourier series as:

\begin{eqnarray}
H_{E}^{\lambda _{\nu }} &=&-\sum_{k}\biggl\{2(\lambda _{\nu }-\cos K)C_{k}^{+}C_{k}+i\gamma
\sin K\left( C_{-k}^{+}C_{k}^{+}+C_{-k}C_{k}\right)  \\
&&+4D\sin KC_{k}^{+}C_{k}-\lambda _{\nu }\biggr\},
\end{eqnarray}
where
\begin{eqnarray}
\begin{split}
&C_{l} =\frac{1}{\sqrt{N}}\sum_{k}e^{iKl}C_{k}\text{ ,} \qquad
C_{l}^{+} =\frac{1}{\sqrt{N}}\sum_{k}e^{-iKl}C_{k}^{+}\text{ ,}\\&
e^{iKN} =1,  \qquad
\frac{1}{N}\sum_{k}e^{i(K\mp R)l} =\delta _{K,\pm R} ,\quad
K =\frac{2\pi k}{N}, \\&
C_{N+1} =C_{1}\text{ ,} \qquad
K =-M,......,-1,0,1,.....,M, \qquad\text{and} \
M =\frac{N-1}{2}.
\end{split}
\end{eqnarray}

The final diagonalized Hamiltonian  $%
H_{E}^{\lambda _{\nu }}$ by  Bogoliubov transformations is expressed by;
\begin{eqnarray}
	H_{E}^{\lambda _{\nu }} =\sum_{k}\Omega _{k}^{\lambda _{\nu }}\left( b_{k}^{+}b_{k}-%
	\frac{1}{2}\right)
	\end{eqnarray}
where,$	\Omega _{k}^{\lambda _{\nu }} =2\sqrt{(\lambda _{\nu }-\cos K)^{2}+\gamma^{2}\sin ^{2}K}+4D\sin K$, and $b_{k}^{+}$ ($b_{k}$) is the  creation (annihilation) operator. The relation between $b_{k}^{+}$ ($b_{k}$) and $C_{l}^{+}$ ($C_{l}$) is given by;
\begin{equation}
	\begin{split}
		&C_{k} =\cos \frac{\theta _{k}^{\lambda _{\nu }}}{2}b_{k,\lambda _{\nu
		}}+i\sin \frac{\theta _{k}^{\lambda _{\nu }}}{2}b_{-k,\lambda _{\nu }}^{+}%
		\text{ ,} \quad
		C_{k}^{+} =\cos \frac{\theta _{k}^{\lambda _{\nu }}}{2}b_{k,\lambda _{\nu
		}}^{+}-i\sin \frac{\theta _{k}^{\lambda _{\nu }}}{2}b_{-k,\lambda _{\nu }},
		\\&
		\text{ and}\quad \theta _{k}^{\lambda _{\nu }} =\tan ^{-1}(\frac{\gamma \sin K}{\lambda
			_{\nu }-\cos K}),\nu =1,2,3,4,
	\end{split}
	\end{equation}
where we choose $\theta _{k}^{\lambda _{\nu }}$ such that the coefficients of all "anomalous",
$b_{k}b_{-k}$ and $b_{k}^{+}b_{k}$, in the energy Hamiltonian $H_{E}^{\lambda _{\nu }}$ vanish. The initial density state of the total system can be expressed as $\rho(0)=\rho _{AB}(0)\otimes \left\vert G\right\rangle \left\langle G\right\vert ,$ where
\begin{equation}
\rho _{AB}(0)=\sum_{\alpha ,\beta }c_{\alpha \beta }\left\vert \chi
_{\alpha }\right\rangle \left\langle \chi _{\beta }\right\vert ,\quad
\alpha ,\beta =1,2,3,4
\end{equation}
also, the state $\left\vert G\right\rangle $ is the ground state of the energy Hamiltonian $H_{E}^{\lambda _{\nu }}$, which is defined as;
\begin{equation}
\left\vert G\right\rangle _{\lambda }=\prod_{k=1}^{M}\cos \frac{\theta _{k}^{\lambda }}{2}\left\vert 0\right\rangle _{k}\left\vert
0\right\rangle _{-k}+i\sin \frac{\theta _{k}^{\lambda }}{2}\left\vert
1\right\rangle _{k}\left\vert 1\right\rangle _{-k}.
\end{equation}
Meanwhile, \begin{eqnarray}
\rho _{AB}(t) &=&Tr[U_{t}\rho _{AB}(0)\otimes \left\vert G\right\rangle
\left\langle G\right\vert U_{t}^{\dagger }] \\
&=&\sum_{\alpha ,\beta }c_{\alpha \beta }S_{\alpha \beta }\left\vert
\chi _{\alpha }\right\rangle \left\langle \chi _{\beta }\right\vert ,
\nonumber \\
S_{\alpha \beta } &=&\left\langle G\right\vert \exp (iH_{T}^{\lambda _{\beta
}}t)\exp (-iH_{T}^{\lambda _{\alpha }}t)\left\vert G\right\rangle ,
\nonumber \\
U_{t} &=&\exp (-iH_{T}^{\lambda _{\nu }}t).
\end{eqnarray}
Hence,
\begin{eqnarray}
S_{\alpha \beta }=\prod_{k>0}&&\biggl\{\cos (\eta _{k}^{\lambda _{\alpha }}-\eta
_{k}^{\lambda _{\beta }})\bigg( \cos \eta _{k}^{\lambda _{\alpha }}\cos \eta
_{k}^{\lambda _{\beta }}e^{\frac{it}{2}(\Omega _{k}^{\lambda _{\alpha
	}}-\Omega _{k}^{\lambda _{\beta }})}+\sin \eta _{k}^{\lambda _{\alpha }}\sin \eta _{k}^{\lambda _{\beta }}e^{-\frac{it}{2}(\Omega _{k}^{\lambda _{\alpha }}-\Omega _{k}^{\lambda _{\beta }})}\bigg) \nonumber \\
&&-\sin
(\eta _{k}^{\lambda _{\alpha }}-\eta _{k}^{\lambda _{\beta }})\bigg(\cos \eta _{k}^{\lambda _{\alpha }}\sin \eta _{k}^{\lambda _{\beta }}e^{\frac{it}{2}%
	(\Omega _{k}^{\lambda _{\alpha }}+\Omega _{k}^{\lambda _{\beta }})}-\sin \eta _{k}^{\lambda _{\alpha }}\cos \eta _{k}^{\lambda _{\beta }}e^{-\frac{it}{2
	}(\Omega _{k}^{\lambda _{\alpha }}+\Omega _{k}^{\lambda _{\beta }})}\bigg)
\biggl\}.
\end{eqnarray}
where, $ \eta _{k}^{\lambda _{j}}=\frac{\theta _{k}^{\lambda _{j}}}{2}-\frac{\theta
	_{k}^{\lambda }}{2}$, is the difference angle between the normal mode dressed by the system-environment interaction and the purely environment.

 \section{Pure state  }
 Let us assume that, the initial system is initially prepared in a generic pure state \cite{Englert2000}.
  By means of the Bloch vectors and the cross dyadic, this pure state can be written as,
	\begin{equation}
		\rho_p(0)=\frac{1}{4}\big( I+ p (\sigma_x-\tau_x)-\sigma_x\tau_x-q(\sigma_y\tau_y+\sigma_z\tau_z)\big)
	\end{equation}
	with $q=\sqrt{1-p^2}$.
The time evolution of this initial state is given by,
\begin{eqnarray}
\rho^P _{AB}(t) &=&\frac{1}{4}\biggl\{(1-q)\big( \left\vert
00\right\rangle \left\langle 00\right\vert +\left\vert 11\right\rangle
\left\langle 11\right\vert \big) +(1+q)\big( \left\vert 01\right\rangle
\left\langle 01\right\vert +\left\vert 10\right\rangle \left\langle
10\right\vert \big) \\
&& +\left( q-1)S_{14}(t)\left\vert 00\right\rangle \left\langle
11\right\vert -(1+q)\left\vert 01\right\rangle \left\langle
10\right\vert  +H.C.\right)\biggl\},  \nonumber
\end{eqnarray}
where $q=\sqrt{1-p^2}$. The eigenvectors  of the initial pure state is obtained as,
	\begin{equation}
	\begin{split}
	& \mu_1(0)=\left\{\frac{1}{\sqrt{2}},0,0,\frac{1}{\sqrt{2}}\right\},\qquad \mu_2(0)=\left\{0,\frac{1}{\sqrt{2}},\frac{1}{\sqrt{2}},0\right\}, \\&
	\mu_3(0)=\left\{-\frac{1}{\sqrt{\frac{2 \text{X1}}{\left| p\right| ^2}+2}},\frac{\left| p\right|  \left(q-\sqrt{p^2+q^2}\right)}{\sqrt{2} p \sqrt{\left| p\right| ^2+\text{X1}}},\frac{\left| p\right|  \left(\sqrt{p^2+q^2}-q\right)}{\sqrt{2} p \sqrt{\left| p\right| ^2+\text{X1}}},\frac{1}{\sqrt{\frac{2 \text{X1}}{\left| p\right| ^2}+2}}\right\},\\&
	 \mu_4(0)=\left\{-\frac{1}{\sqrt{\frac{2 \text{X1}}{\left| p\right| ^2}+2}},\frac{\left| p\right|  \left(\sqrt{p^2+q^2}+q\right)}{\sqrt{2} p \sqrt{\left| p\right| ^2+\text{X1}}},-\frac{\left| p\right|  \left(\sqrt{p^2+q^2}+q\right)}{\sqrt{2} p \sqrt{\left| p\right| ^2+\text{X1}}},\frac{1}{\sqrt{\frac{2 \text{X1}}{\left| p\right| ^2}+2}}\right\};
	\end{split}
	\end{equation}
	where, $X_1=| 2 q \left(\sqrt{p^2+q^2}+q\right)+p^2| $.
	The eigenvectors of the final state is given by:
	\begin{equation}
	\begin{split}
	&\psi_1(t)=\left\{0,\frac{1}{\sqrt{2}},\frac{1}{\sqrt{2}},0\right\},\quad \psi_2(t)=\left\{0,-\frac{1}{\sqrt{2}},\frac{1}{\sqrt{2}},0\right\}\\&
	\psi_3(t)=\left\{-\frac{| S(1,4)t| }{S(4,1) \sqrt{| \frac{S(1,4)}{S(4,1)}| +1}},0,0,\frac{1}{\sqrt{| \frac{S(1,4)}{S(4,1)}| +1}}\right\},\\&
	\psi_4(t)=\left\{\frac{| S(1,4)t| }{S(4,1) \sqrt{| \frac{S(1,4)}{S(4,1)}| +1}},0,0,\frac{1}{\sqrt{| \frac{S(1,4)}{S(4,1)}| +1}}\right\}.
	\end{split}
	\end{equation}

\section{Dynamics of orthogonality}
The aim of this section is shedding the light on the  orthogonality  behavior of the final  state of the qubits system, where it is assumed that the two qubit system is initial prepared in a maximum entangled state (MES) or partial entangled state (PES). In this context, it is important do define the concept of the orthogonality.  Now, consider that a system is initially prepared in the state $\psi(0)$, this state is evolved under the unitary operator $\mathcal{U}(t)=e^{-i\mathcal{H}t}$, where $\mathcal{H}$ is the Hamiltonian which describes the system. Then, the final state $\ket{\psi(t)}=\mathcal{U}(t)\ket{\psi(0)}$. The orthogonality is given by \cite{margolus1998maximum,metwally2012information,obada2011quantum}:

\begin{equation}
\mathcal{S}_{or}=\expect{\psi(t)|\psi(0)}.
\end{equation}
In our treatment, we shall investigate the orthogonality of the eigenvectors  of the initial and the final density operators. In our numerical calculations, it is enough to consider only two components of the eigenvectors to  examine the effect of the parameters of the field and the initial state settings.

\subsection{\it The system is initially prepared in a partial pure state}

 \begin{figure}[h!]
	\includegraphics[width=0.5\linewidth, height=6cm]{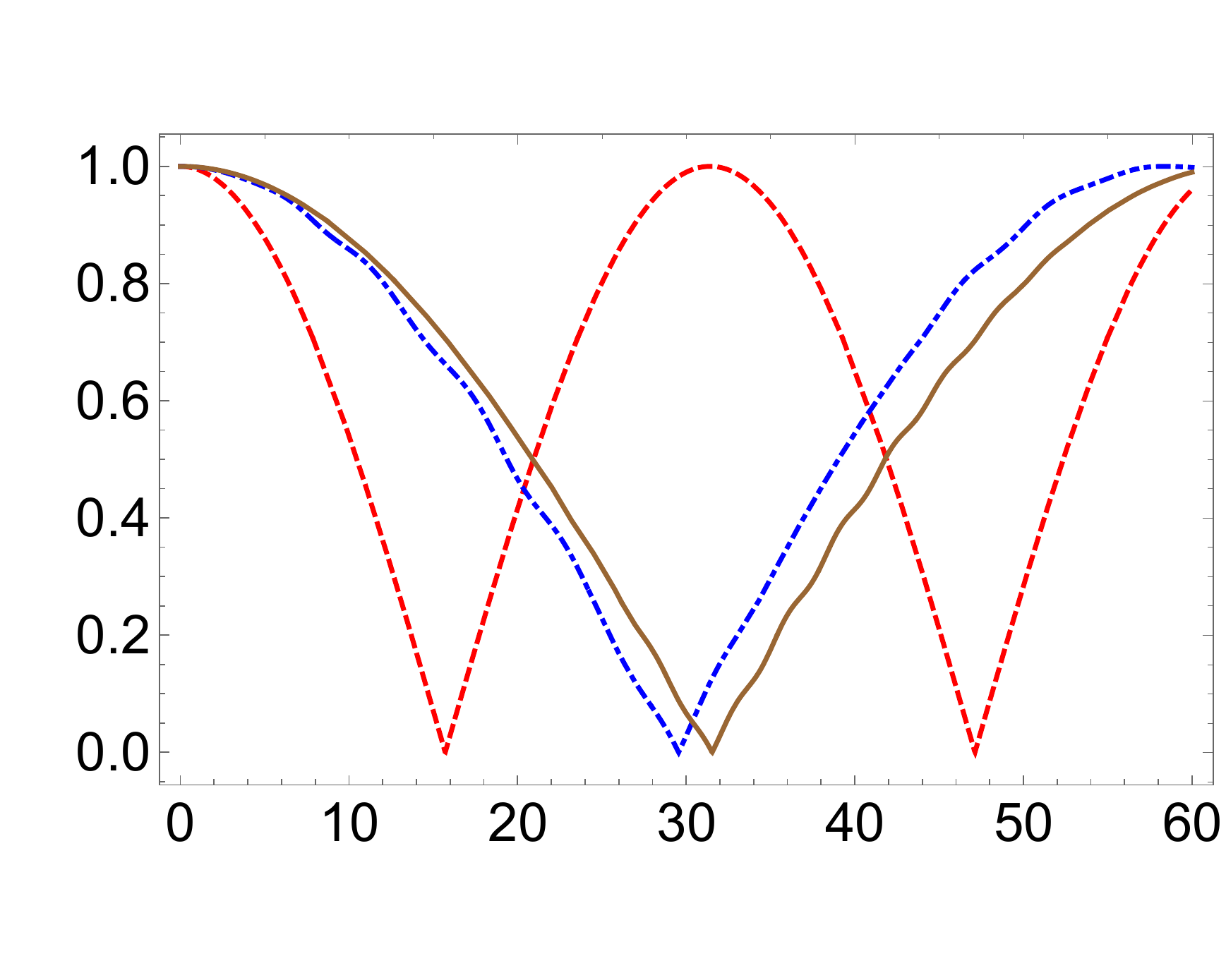}
\put(-110,150){$(a)$}
\put(-240,90){$\mathcal{S}_{or}$}
\put(-110,10){$t$}
\includegraphics[width=0.5\linewidth, height=6cm]{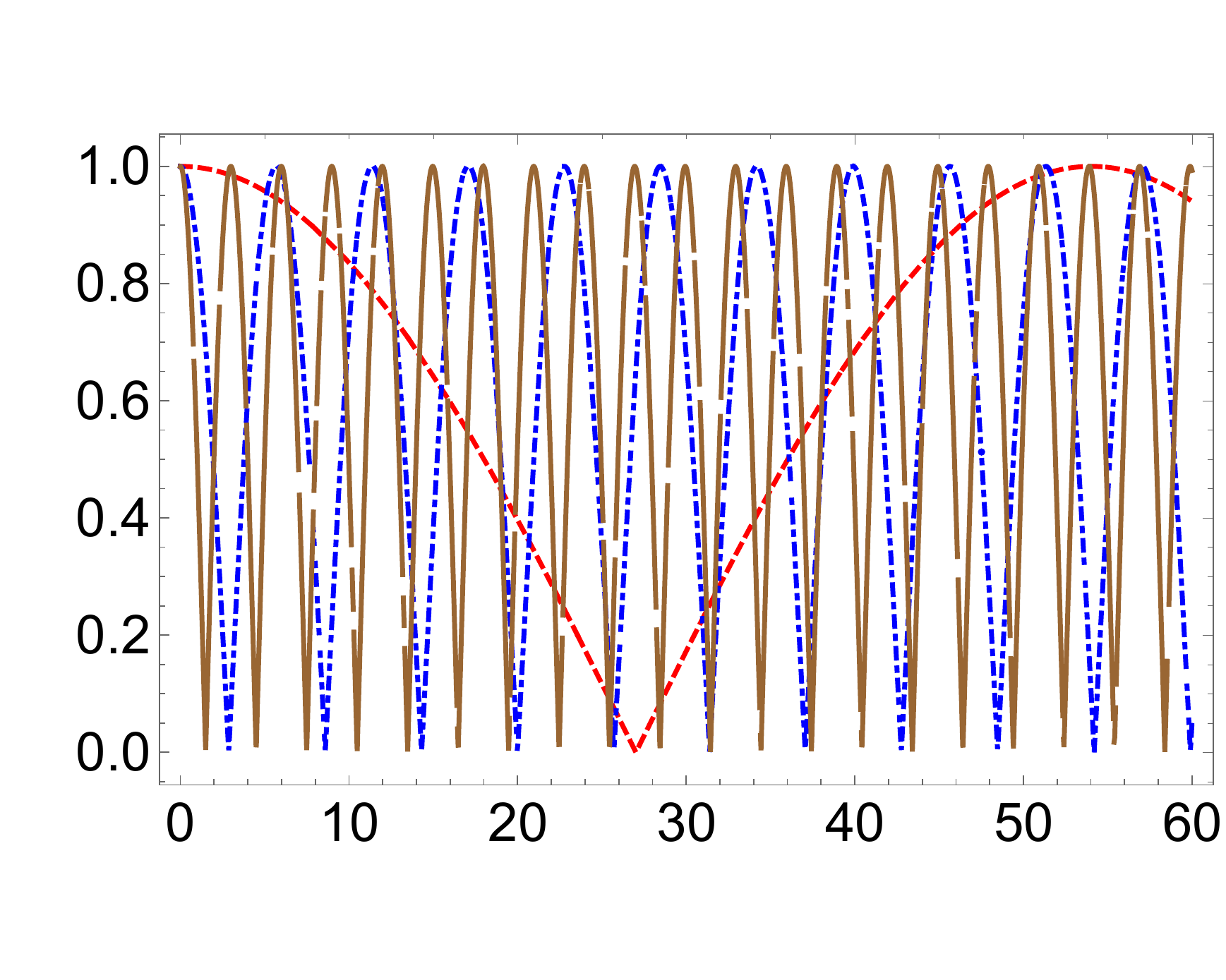}
\put(-110,150){$(b)$}
\put(-235,90){$\mathcal{S}_{or}$}
\put(-110,10){$t$}
	\caption{ The orthogonality speed of the qubit system where the two qubit system interacts with environment consists of N sites, $7,27$ for (a),(b), respectively, where  $D=0$ ,g=0.1 $\lambda=0$, p=0.5.}
	\label{g.2}
\end{figure}
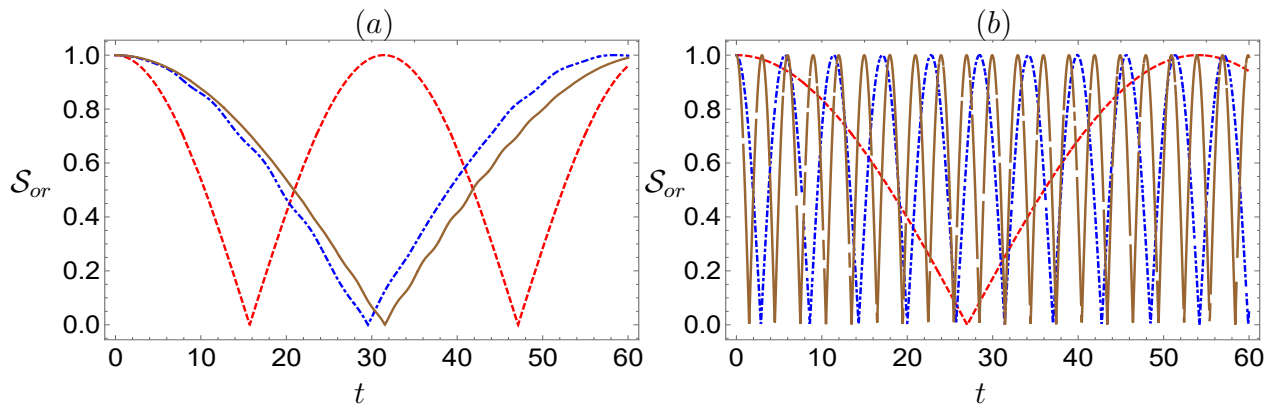

In Fig.(\ref{g.2}), we discuss the speed of orthogonality of a system prepared in a pure state Consisted of two qubits. This qubit system  interacts locally with environment consists of $N$ sites of $7$ and $27$ qubits.  Moreover, we  consider that the interaction system is prepared in the $XX$ chain model, i.e., $\gamma=0$, in anisotropic case,$0<\gamma<1$, and the Ising model with $\gamma=1$.  It is clear that, the speed of orthogonality $\mathcal{S}_{or}$ depends on the sites number, where as one increases the site number $N$, the number of oscillations increasing and the orthogonality is displayed at short time. This means that the possibility of transmitting the information increases as $N$ increases.  On the other hand, the type of the interaction model plays an important role on decreasing the time of orthogonality. However, as it is displayed from Fig.(\ref{g.2}a) the shortest orthogonal time is depicted for the $XX$- chain model $(\gamma=0)$, while the largest one is displayed  for Ising model.  These results are coincide with the definition of the  orthogonal time, which is defined $T\geq \frac{\bar h}{4E}$, where $E$ is the energy, where $E_{\gamma=0}>E_{\gamma=0.5}>E_{\gamma=1}$.  However, this results will be changed  depending on the site numbers $N$. Therefore as $N$ increase $E_{\gamma=1}$ will be the largest and consequently the orthogonal time will be the smallest.

\begin{figure}[h!]
	\includegraphics[width=0.5\linewidth, height=6cm]{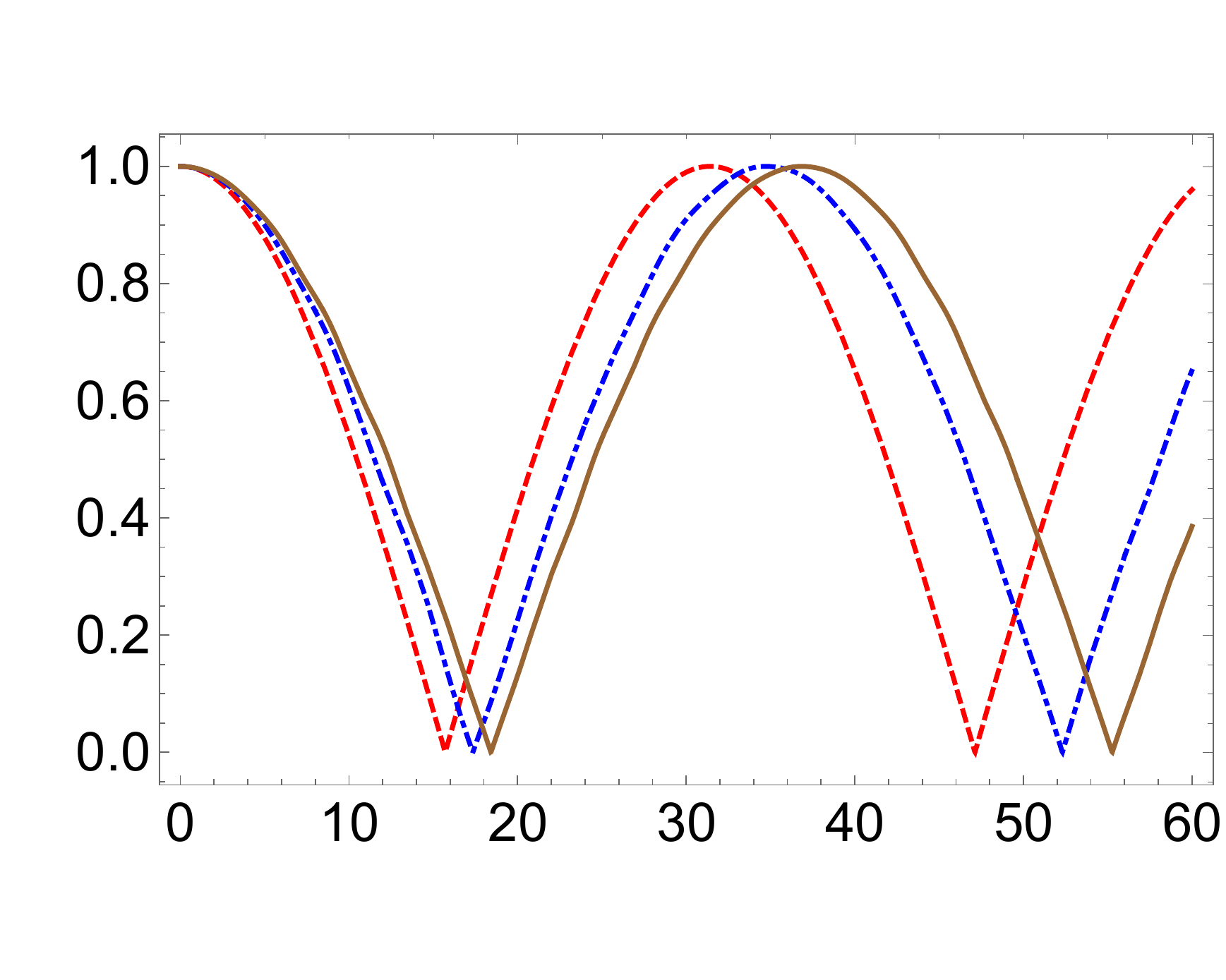}~~\quad
	\put(-110,150){$(a)$}
\put(-240,90){$\mathcal{S}_{or}$}
\put(-110,10){$t$}
	\includegraphics[width=0.5\linewidth, height=6cm]{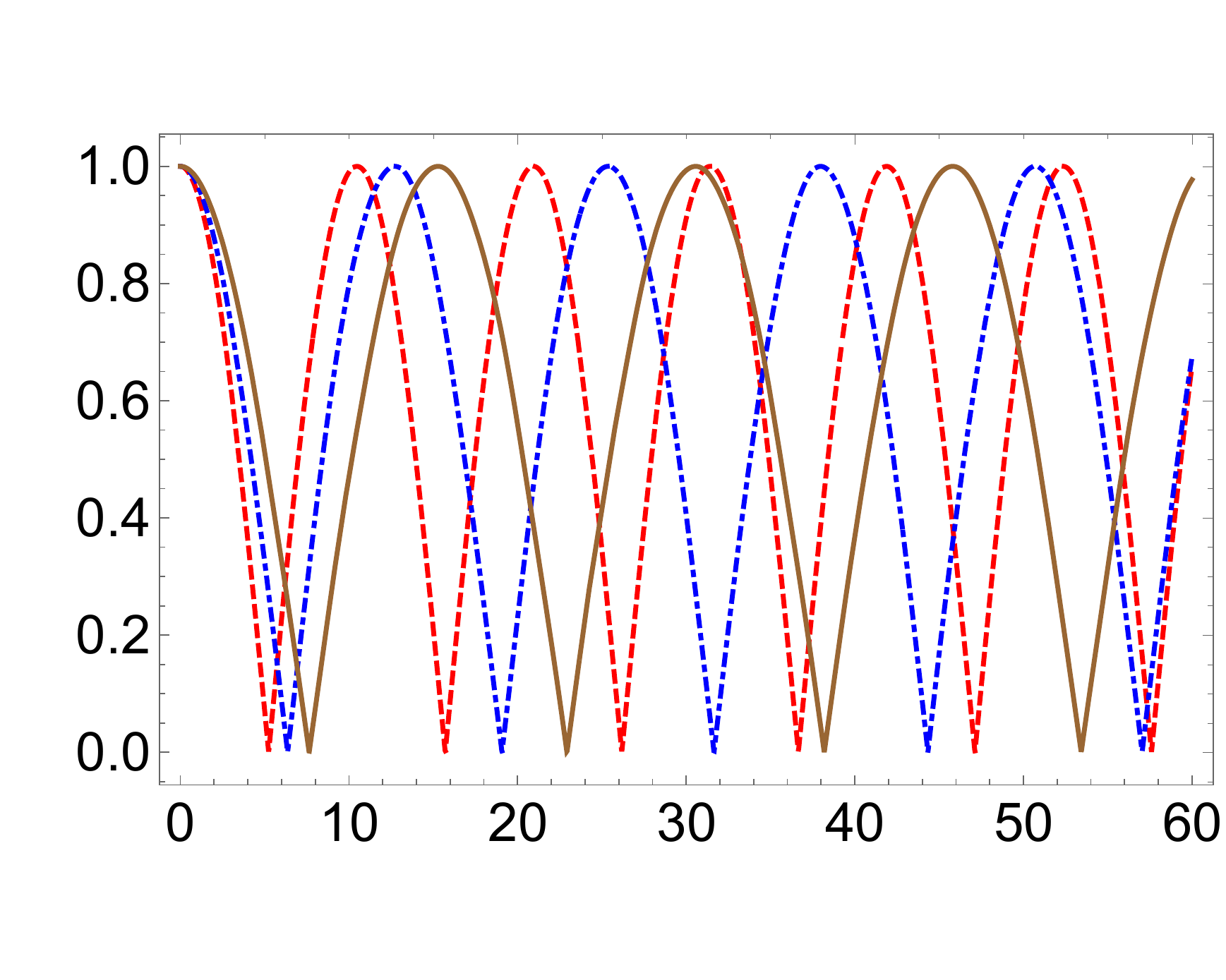}~\quad
	\put(-110,150){$(b)$}
\put(-235,90){$\mathcal{S}_{or}$}
\put(-110,10){$t$}
	\caption{The orthogonality speed $\mathcal{S}_{or}$, where the initial state is prepared with $p=0.5$ and
 $N=7$, $D=0, g=0.1$ and (a)$\lambda=0.2$, (b) $\lambda=0.9$.}
	\label{g.3}
\end{figure}
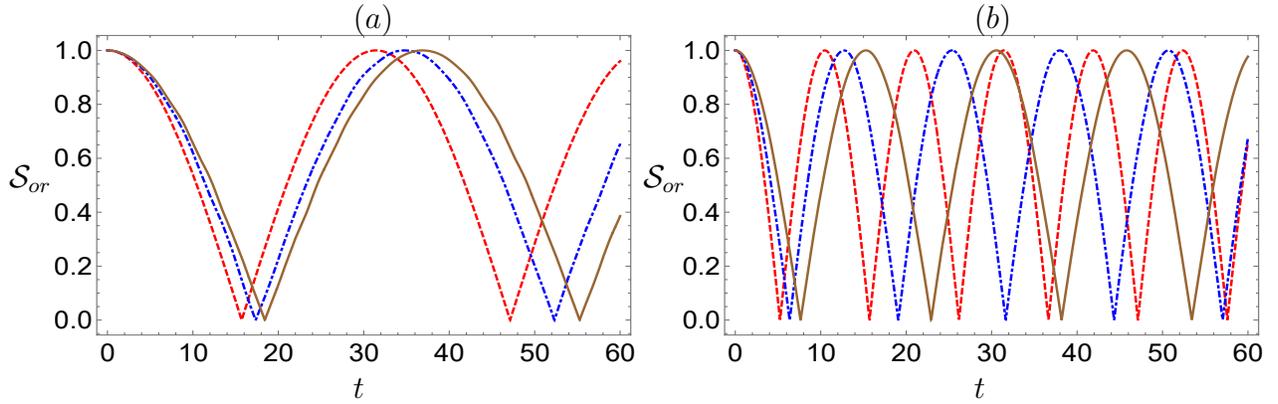

To investigate the effect of $\lambda$; the strength of the external filed, we consider that the number of sites $N=7$
 and the initial pure state is prepared with $p=0.5$.  Fig.(\ref{g.3} displays the behavior of  $\mathcal{S}_{or}$ for the three different types of the interaction,   namely, the $XX$ chain mode, Ising model, and the isotropic model at large strength where we set $\lambda=0.2,0.9$
 It is clear that, the number of orthogonality is larger than those depicted on the absence of the external field.

\begin{figure}[h!]
\centering
	\includegraphics[width=0.5\linewidth, height=6cm]{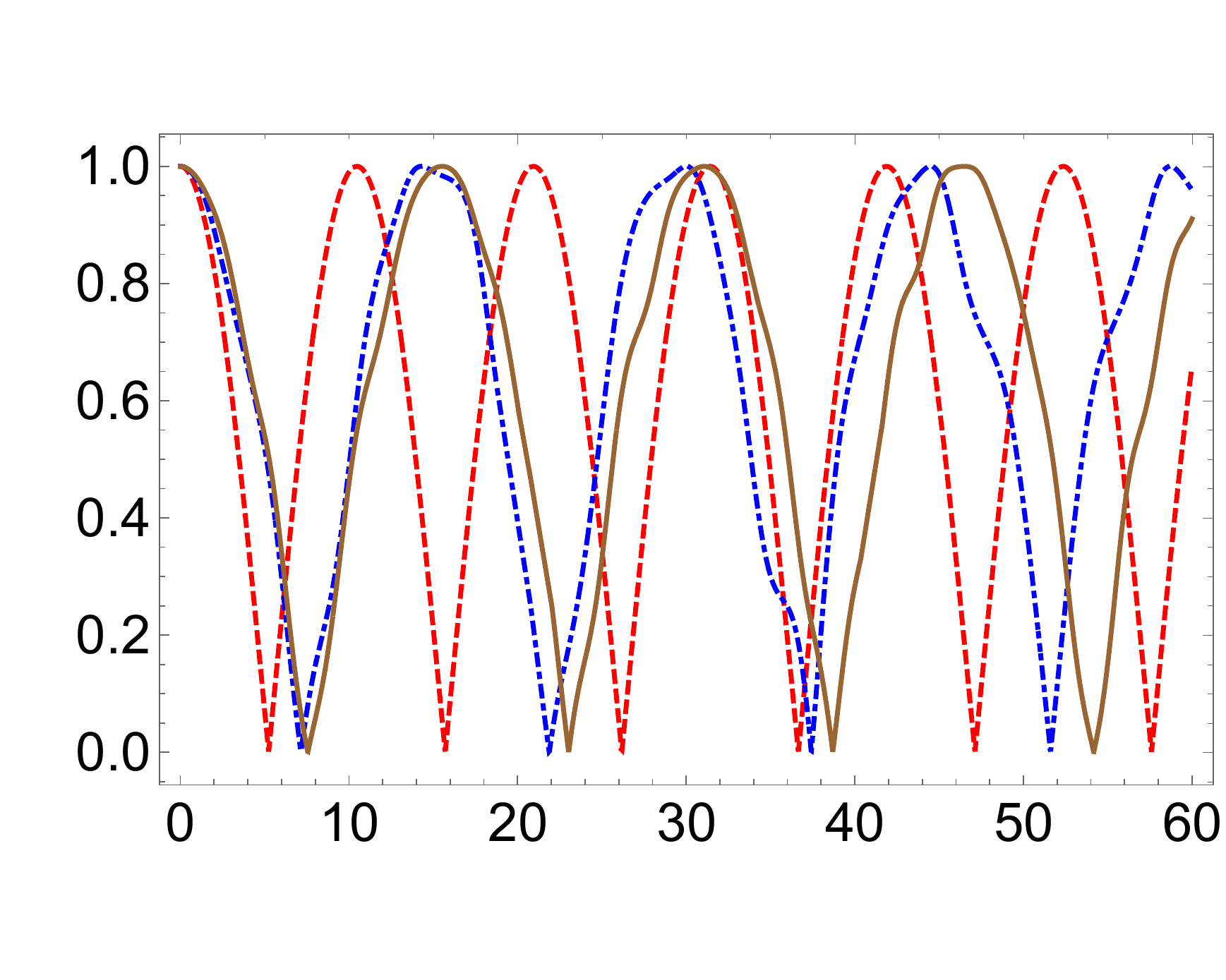}
\put(-240,90){$\mathcal{S}_{or}$}
\put(-110,10){$t$}
	\caption{The effect of the coupling constant  same as Fig(\ref{g.2}a) but we set $g=0.3$}
	\label{g.4}
\end{figure}

The effect of the coupling constant between the two-qubit system and the qubits of the environment is displayed in Fig.(\ref{g.4}), where we set $g=0.3$. It is clear that, the number of orthogonality increases as the coupling constant increases. These result appears clearly by comparing Fig.(\ref{g.2}a) and Fig.(\ref{g.4}), where the first orthogonality is depicted for the $XX$ chain model at short time compared with that displayed in Fig.(\ref{g.3}a). Although the number of orthogonality that depicted for Ising and anisotropic models are the same, but the time of orthogonality that depicted for the anisotropic interaction is smaller than that shown for the Ising model.

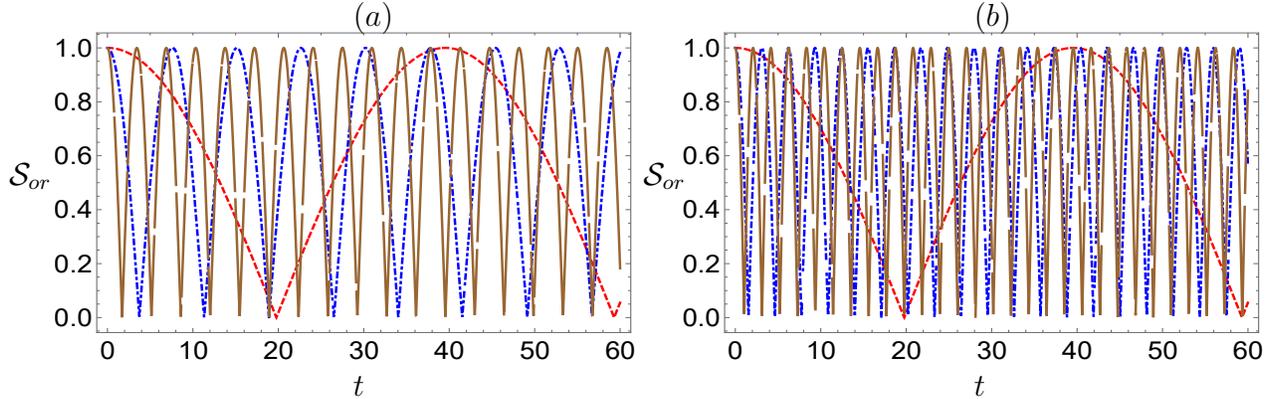
\begin{figure}[h!]
	\includegraphics[width=0.5\linewidth, height=6cm]{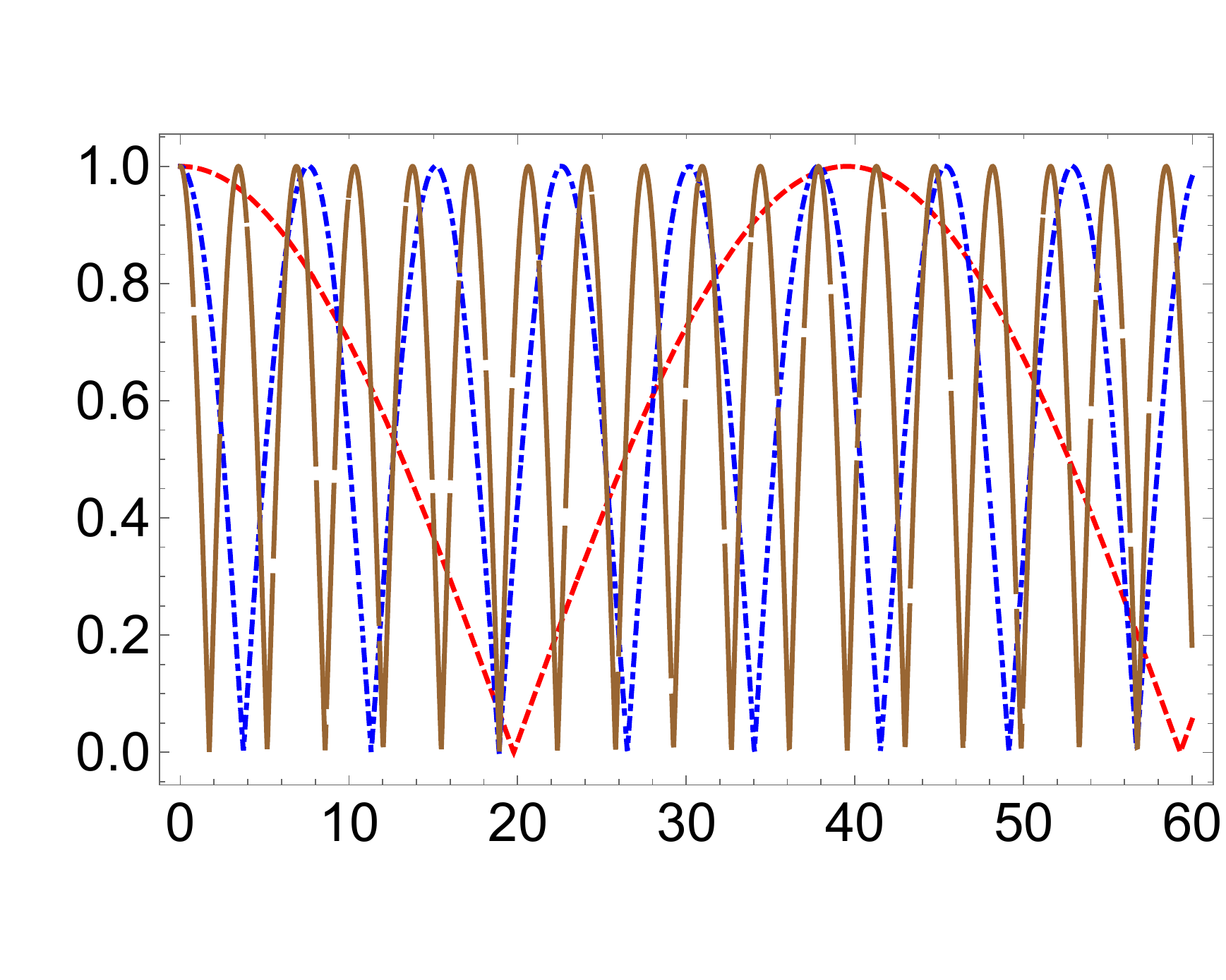}~~\quad
\put(-110,150){$(a)$}
\put(-240,90){$\mathcal{S}_{or}$}
\put(-110,10){$t$}
	\includegraphics[width=0.5\linewidth, height=6cm]{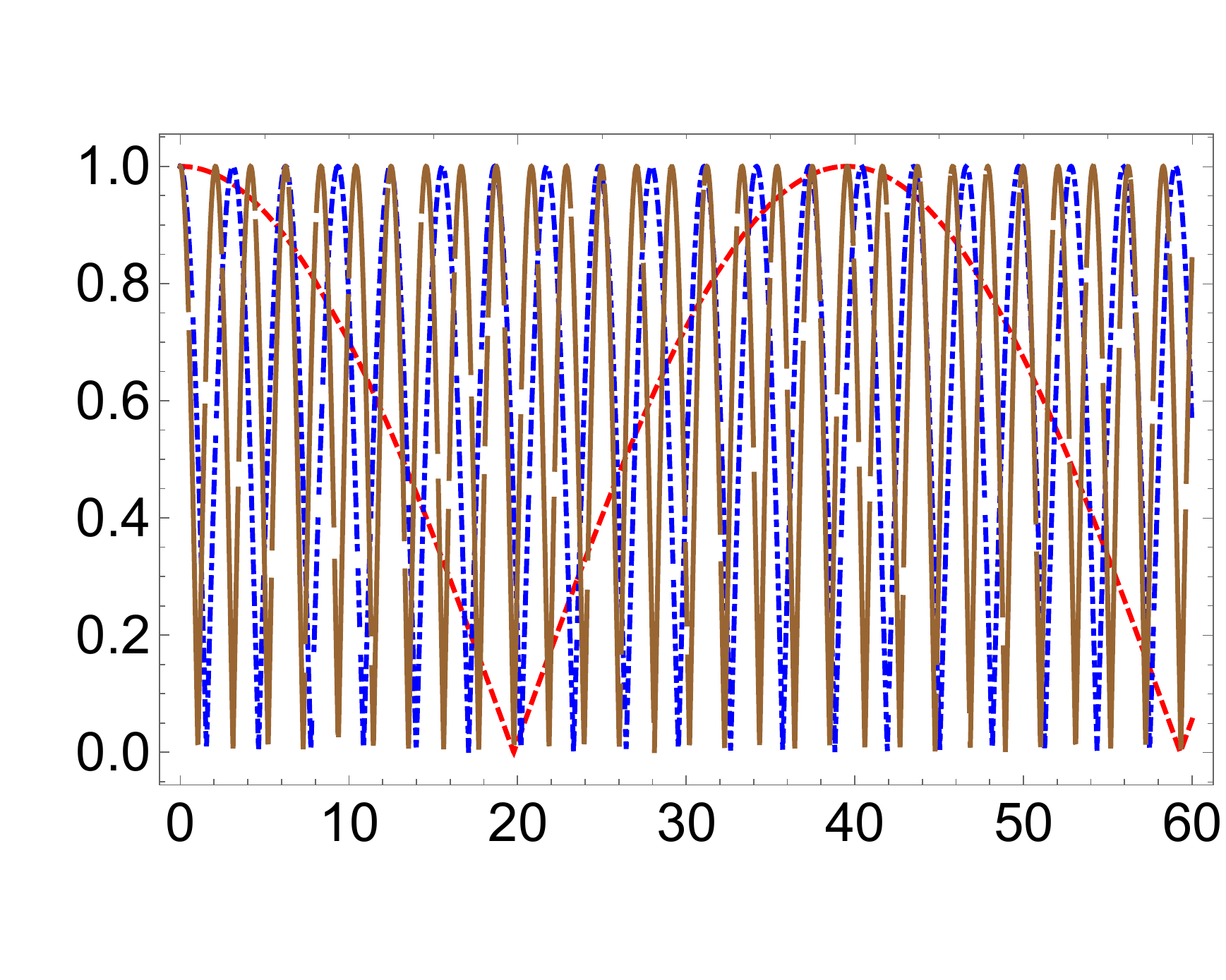}~\quad
	\put(-110,150){$(b)$}
\put(-235,90){$\mathcal{S}_{or}$}
\put(-110,10){$t$}
	\caption{The effect of the DM interaction on  orthogonality speed $\mathcal{S}_{or}$, where the initial state is prepared with $p=0.5$ and
 $N=13, g=0.1$,$\lambda=0.1$, (a) $DM=0$  and (b) $DM=0.3$.}
	\label{g.1}
\end{figure}
In Fig.(\ref{g.1}), we discuss the effect of the $DM$ interaction on the behavior of the orthogonality speed, where we consider that the initial system is initially prepared in a partial entangled pure state with $p=0.5$. In this context, we would like to mention that, the impact effect of $DM$  appears only at large numbers of the environment's qubits and the existence of the external field. As it is displayed from Fig.(\ref{g.1}a), due to the large numbers of the initial environment's qubit, i,e. we set $N=13$, the number of orthogonality increases and consequently the time of orthogonality decreases. In Fig.(\ref{g.1}b), we increase the strength of the Dzyaloshinsky-Moriya interaction, where $DM=0.3$. It is clear that, the behavior of $\mathcal{S}_{or}$ is similar to that displayed in Fig.(\ref{g.1}a), but with larger numbers of orthogonality and shorter time of orthogonality.  Moreover, by  switching the $DM$ interaction, the effectiveness of the interactions types  will  change. However, the impact of Ising model on the orthogonality time is the shortest  one, while the longest  one is depicted for the   $XX$  chain model.
Therefore, to decrease the orthogonality time, and consequently, the computational speed, one has to switch on the  Dzyaloshinsky-Moriya interaction in the presences of  any strength of the external field with large number of environment's qubits.

\begin{figure}[h!]
\centering
	\includegraphics[width=0.4\linewidth, height=5cm]{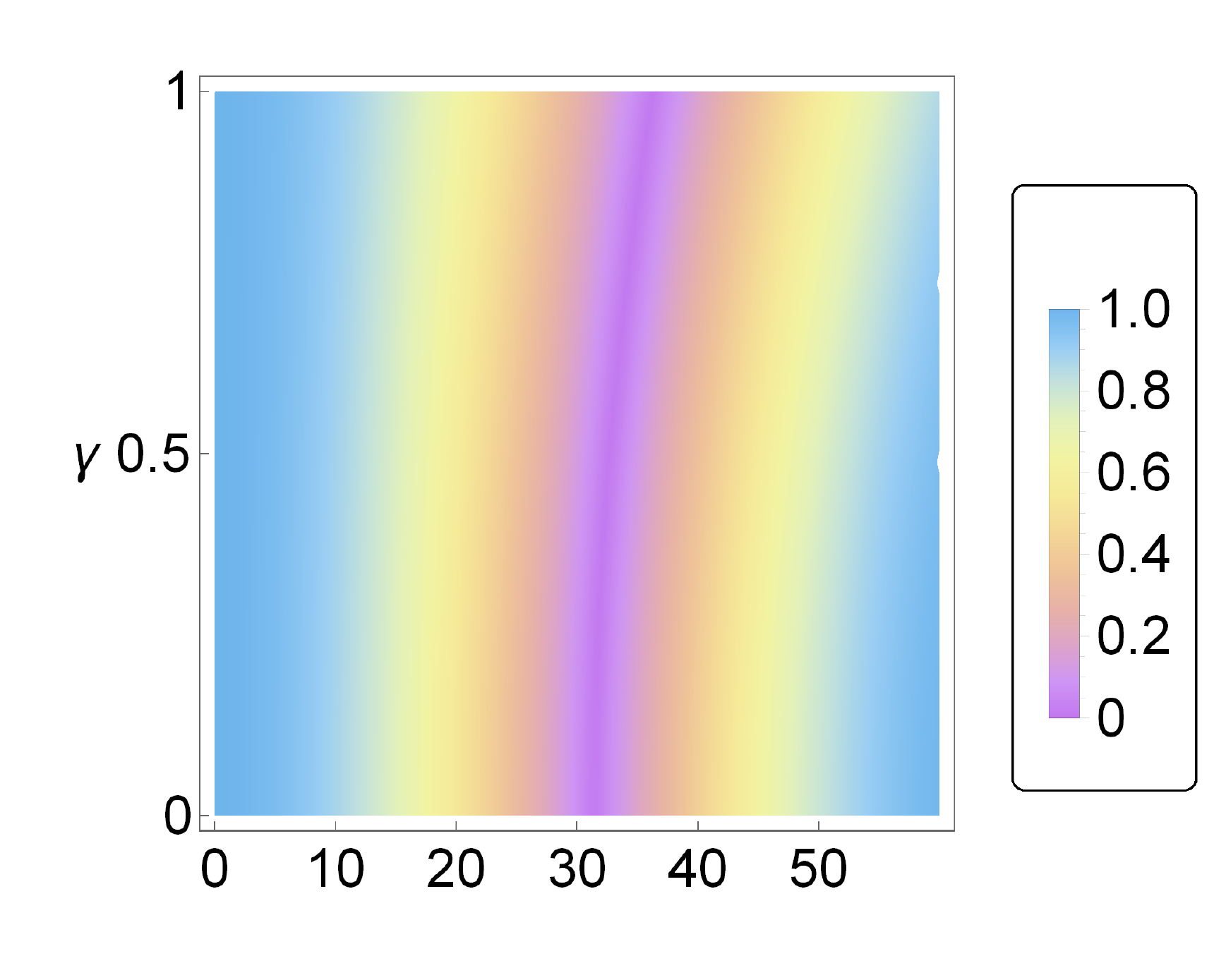}~~\quad
\put(-110,140){$(a)$}
	\includegraphics[width=0.4\linewidth, height=5cm]{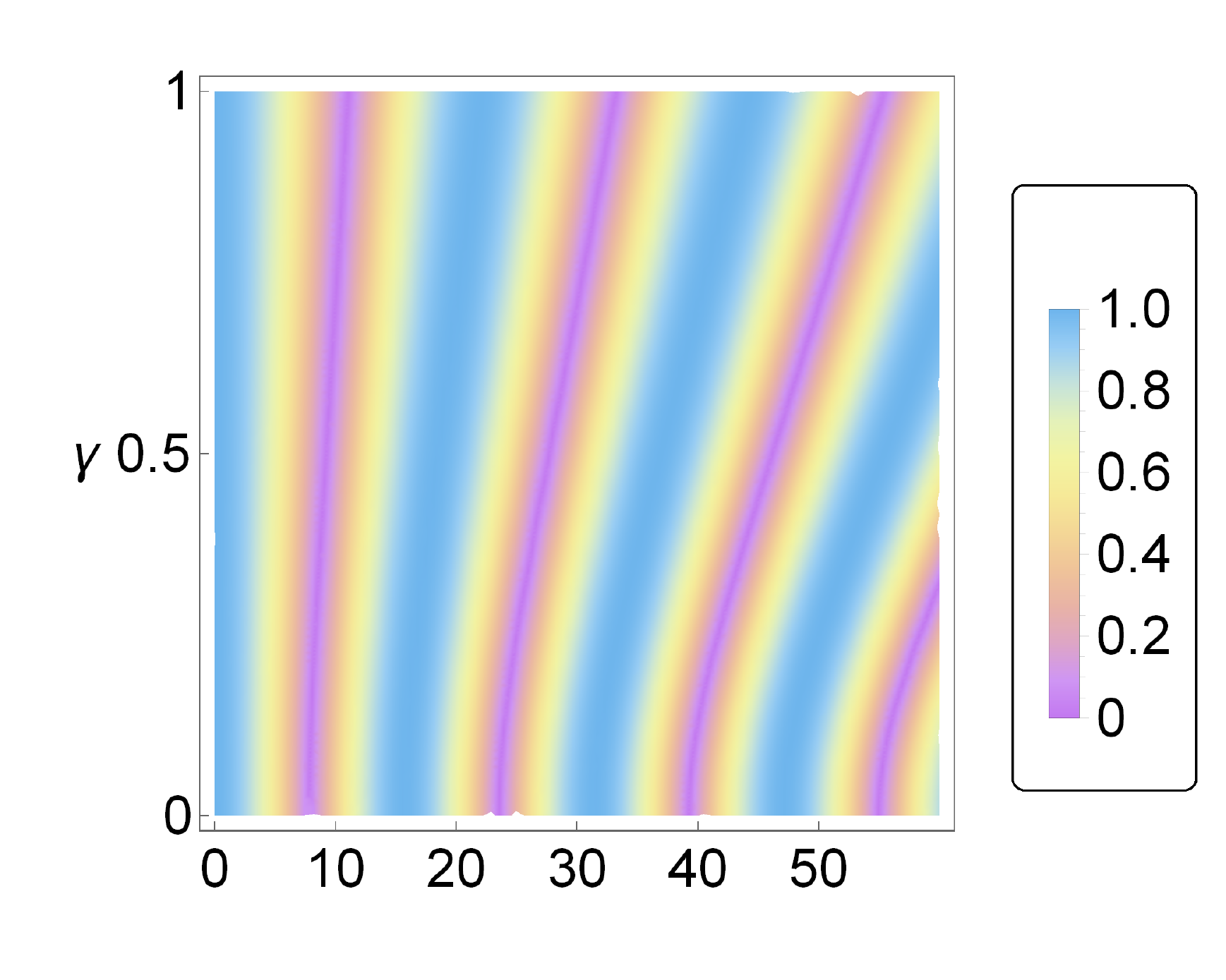}~~\quad
\put(-110,140){$(b)$}
	\caption{The behavior of $\mathcal{S}_{or}(\gamma)$ for a system is initially prepared in a PES ($p=0.5)$ with , $D=0.5$, g=0.05 $ \lambda=1$, where
(a) $N=3$, (b) $N=9$.}
	\label{g.5}
\end{figure}

Fig.(\ref{g.5}), shows the behavior of the orthogonality $\mathcal{S}_{or}$  as contour plot, where it is assumed that the initial environment consists of small number of qubits, $N=3, 9$. It is clear that, at small number ($N=3$, Fig. (\ref{g.5}a)), the first orthogonality for the three types of interactions appears at $t\simeq 30$. However as $\gamma$ increases, namely the interaction turns into anisotropic ($0<\gamma<1$), the orthogonality is depicted at larger time. On the other hand, the largest orthogonal time is displayed  for the Ising model, namely at $\gamma=1$.  As one increases the qubits of the environment  ($N=9$, Fig. (\ref{g.5}b)),  the number of orthogonality increases. Moreover, the time delay of orthogonality is displayed as  $\gamma$ increases.

\subsection{\it The initial system is initial prepared in a MES}

Finally, we  assume that the system is initially  prepared in a maximum entangled  state  of Bell type.
$|\phi^+\rangle = \frac{1}{\sqrt{2}}\big(|00\rangle+|11\rangle\big)$. The eigenvectors for this state is given by;
\begin{equation}
\begin{split}
&\mu_1(0)=\left\{0,1,0,0\right\},\qquad \mu_2(0)=\left\{0,0,1,0\right\}, \\&
\mu_3(0)=\left\{\frac{1}{\sqrt{2}},0,0,\frac{1}{\sqrt{2}}\right\},\qquad \mu_4(0)=\left\{-\frac{1}{\sqrt{2}},0,0,\frac{1}{\sqrt{2}}\right\}
\end{split}
\end{equation}
The time evolution of this state is given by,
\begin{eqnarray}
\rho^m _{AB}(t) &=&\frac{1}{2}\biggl\{ \left\vert
00\right\rangle \left\langle 00\right\vert +\left\vert 11\right\rangle
\left\langle 11\right\vert +S_{14}(t)\left\vert 00\right\rangle \left\langle
11\right\vert +H.C.\biggl\}.  \nonumber
\end{eqnarray}
 The eigenvectors of the final state are defined as,
\begin{equation}
\begin{split}
&\psi_1(t)=\left\{0,1,0,0\right\},\quad \psi_2(t)=\left\{0,0,1,0\right\}\\&
\psi_3(t)=\left\{ \sqrt{\frac{S_{14}(t)}{2 S_{41}(t)}} ,0,0,\frac{1}{\sqrt{2}}\right\},\quad
\psi_4(t)=\left\{ -\sqrt{\frac{S_{14}(t)}{2 S_{41}(t)}} ,0,0,\frac{1}{\sqrt{2}}\right\}
\end{split}
\end{equation}

The  effect of the interaction's parameters on the behavior of the orthogonality speed for a system is initially prepared in a maximum entangled state,
is similar to that displayed for seperabpe/partial entangled states. However, it is important to  clarify our results by examining the behavior of
 $\mathcal{S}_{or}$, where we assume that the system is initially prepared in the state
  $\rho(0)=\ket{\phi^+}\bra{\phi^+}, ~\phi^+=\frac{1}{\sqrt{2}}(\ket{00}+\ket{11})$. In Fig.(\ref{g.8}a), we assume that the environment consists of $3$ qubits and the same parameters are fixed as in Fig. (\ref{g.5})).
    The behavior  of $\mathcal{S}_{or}$ is similar to that displayed in Fig. (\ref{g.5}a)), where the initial system is initially prepared in a partial entangled   state with $p=0.5$.  However, the numbers of orthogonality that displayed in Fig.(\ref{g.8}a) is larger than that displayed in Fig. (\ref{g.5}a)), namely the time of orthogonality is smaller, where the first orthogonality is displayed at $t\simeq 15$. In Fig.(\ref{g.8}b), we increase the qubits of the environment, $(N=9)$. The behavior of $\mathcal{S}_{or}$ shows the number of orthogonality increases, and consequently the time of orthogonality decreases. Moreover, the delay of orthogonality time increases as $\gamma$ increases.

  From  Fig.(\ref{g.5}) and (\ref{g.8}) , one may conclude that preparing the initial  two-qubit system in a maximum entangled state, (MES) increases the number of orthogonality and  as a result the speed of transporting  information  and  the  computations increases.
\begin{figure}[h!]
\centering
	\includegraphics[width=0.4\linewidth, height=5cm]{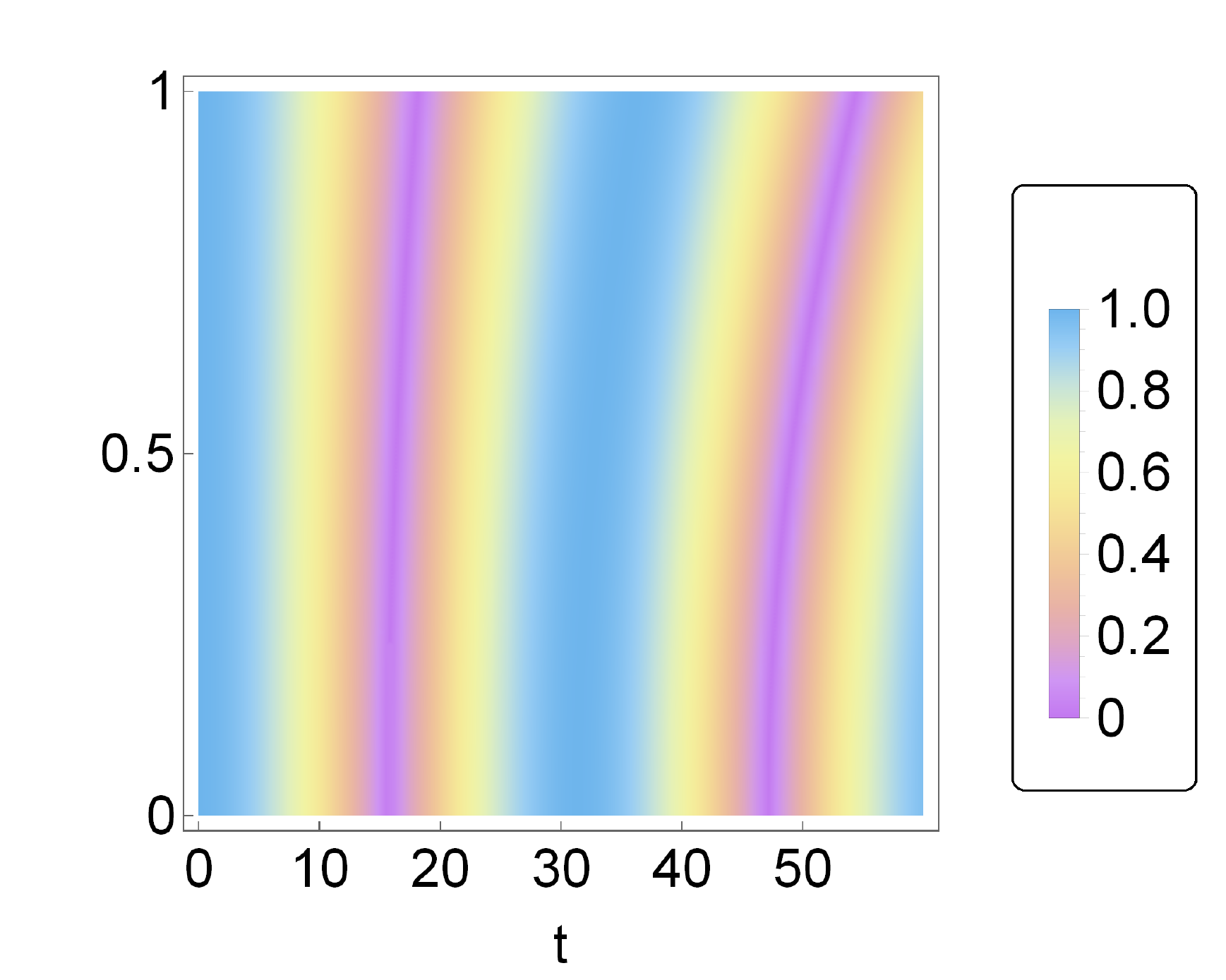}~~\quad
\put(-185,80){$\gamma$}
\put(-110,140){$(a)$}
	\includegraphics[width=0.4\linewidth, height=5cm]{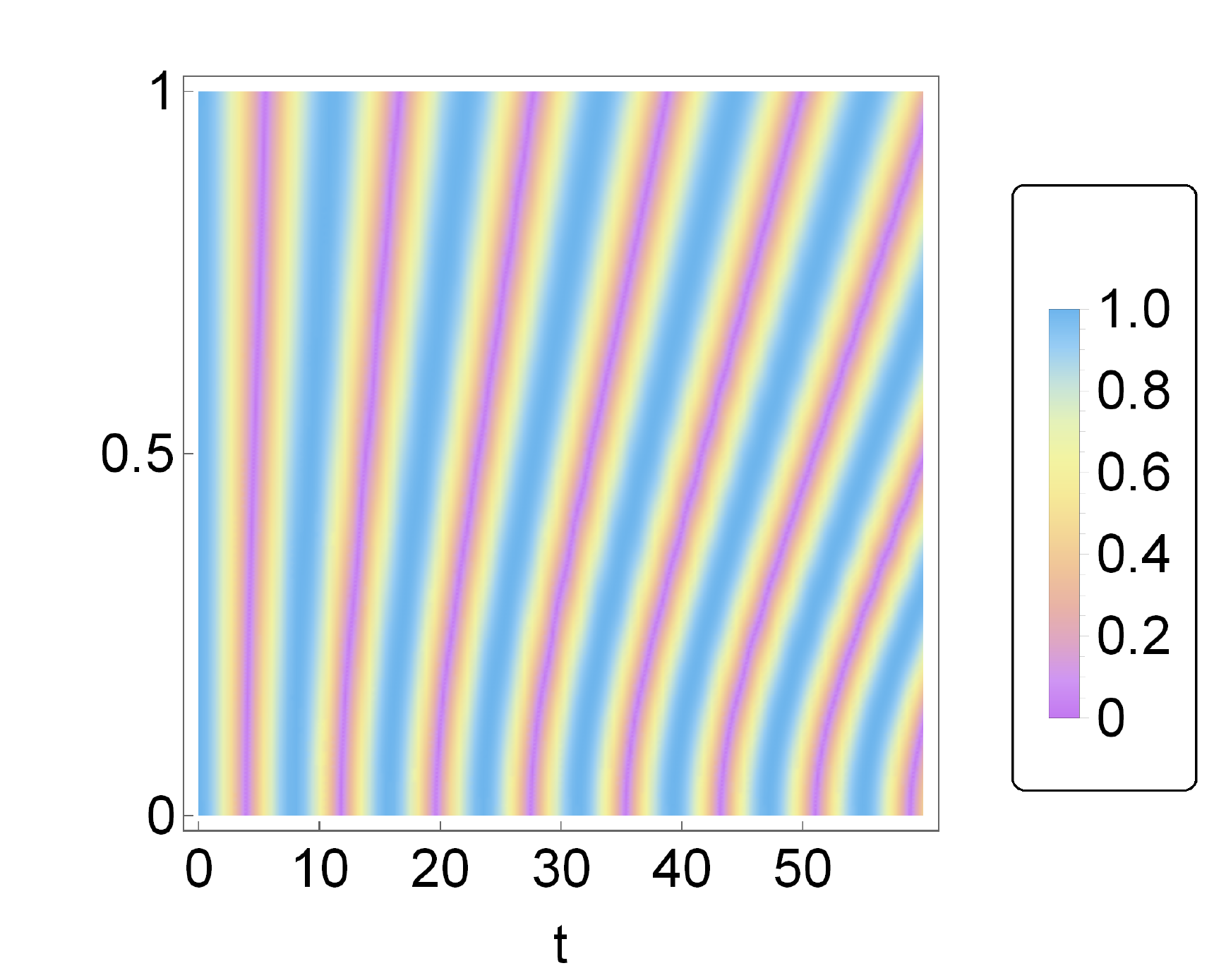}~~\quad
\put(-185,80){$\gamma$}
\put(-110,140){$(b)$}
	\caption{The behavior of $\mathcal{S}_{or}(\gamma)$ for a system is initially prepared in a MES with , $D=0.5$ ,g=0.05 $ \lambda=1$, where
(a) $N=3$, (b) $N=9$.}
		\label{g.8}
\end{figure}
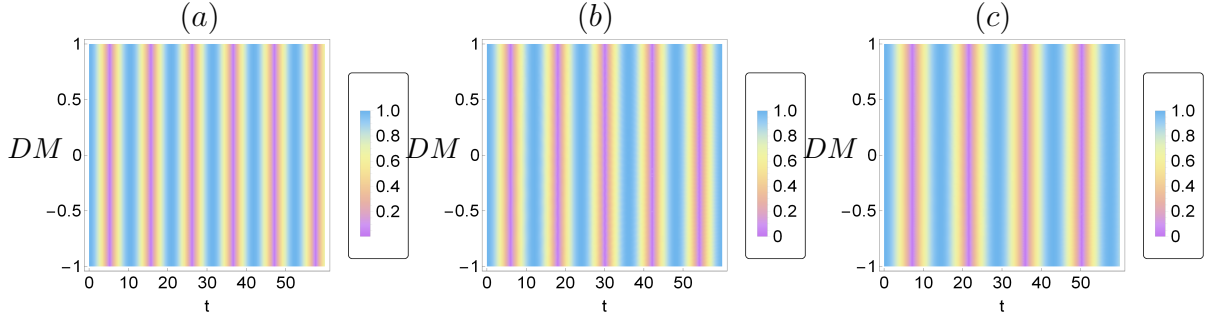
\begin{figure}[t!]
\centering
	\includegraphics[width=0.32\linewidth, height=4cm]{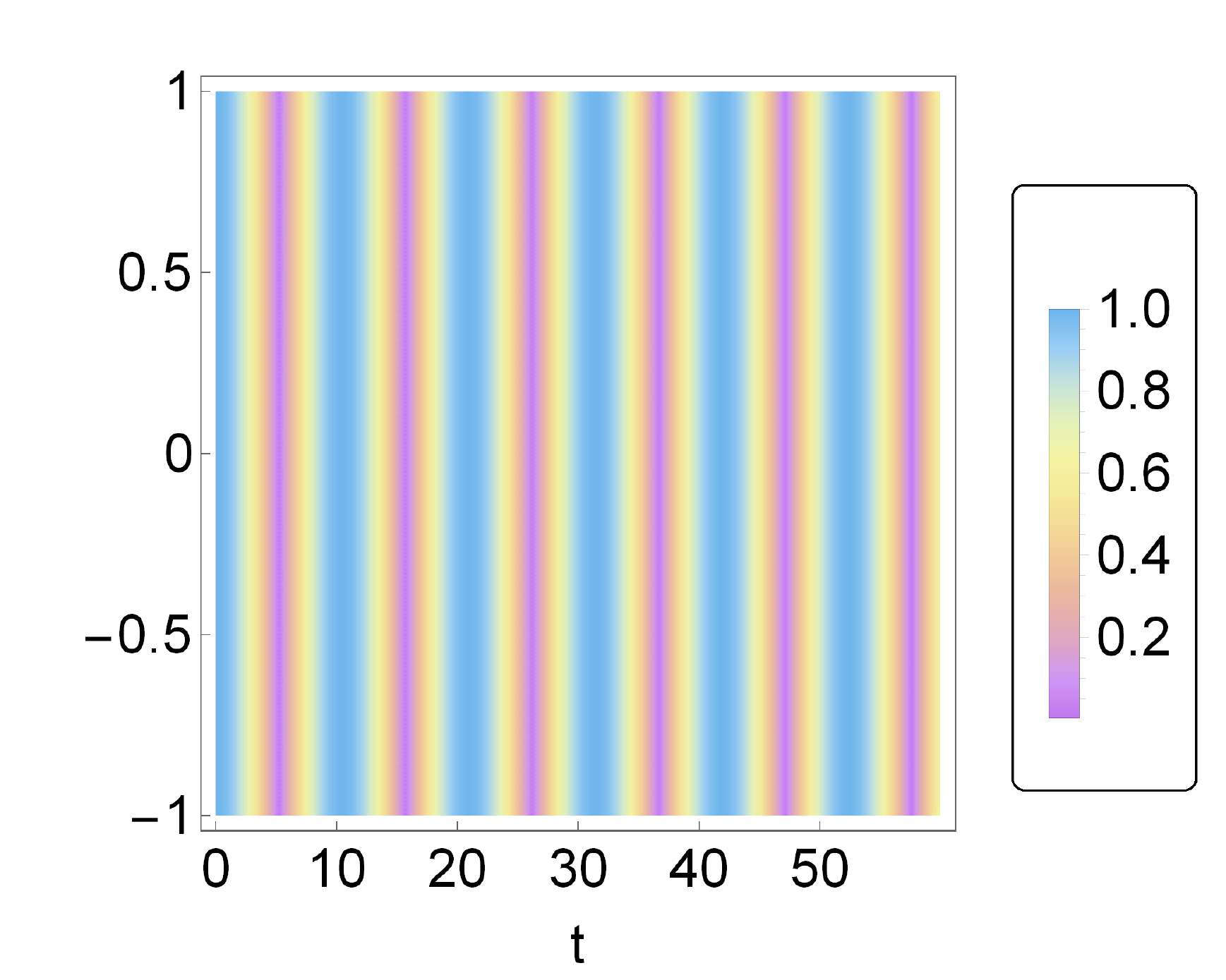}
\put(-155,60){\small$DM$}
\put(-90,110){$(a)$}
	\includegraphics[width=0.32\linewidth, height=4cm]{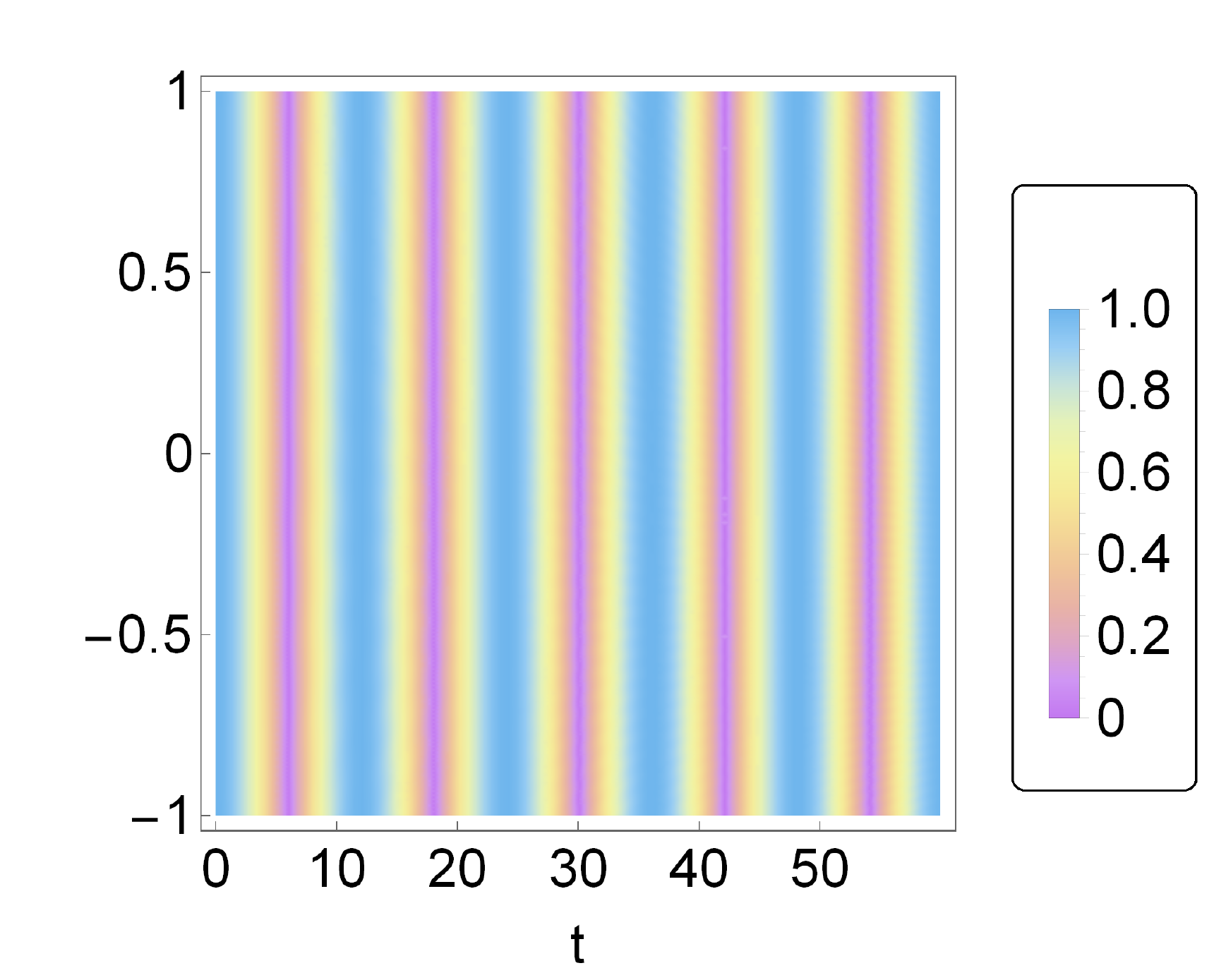}
\put(-155,60){\small $DM$}
\put(-90,110){$(b)$}
	\includegraphics[width=0.32\linewidth, height=4cm]{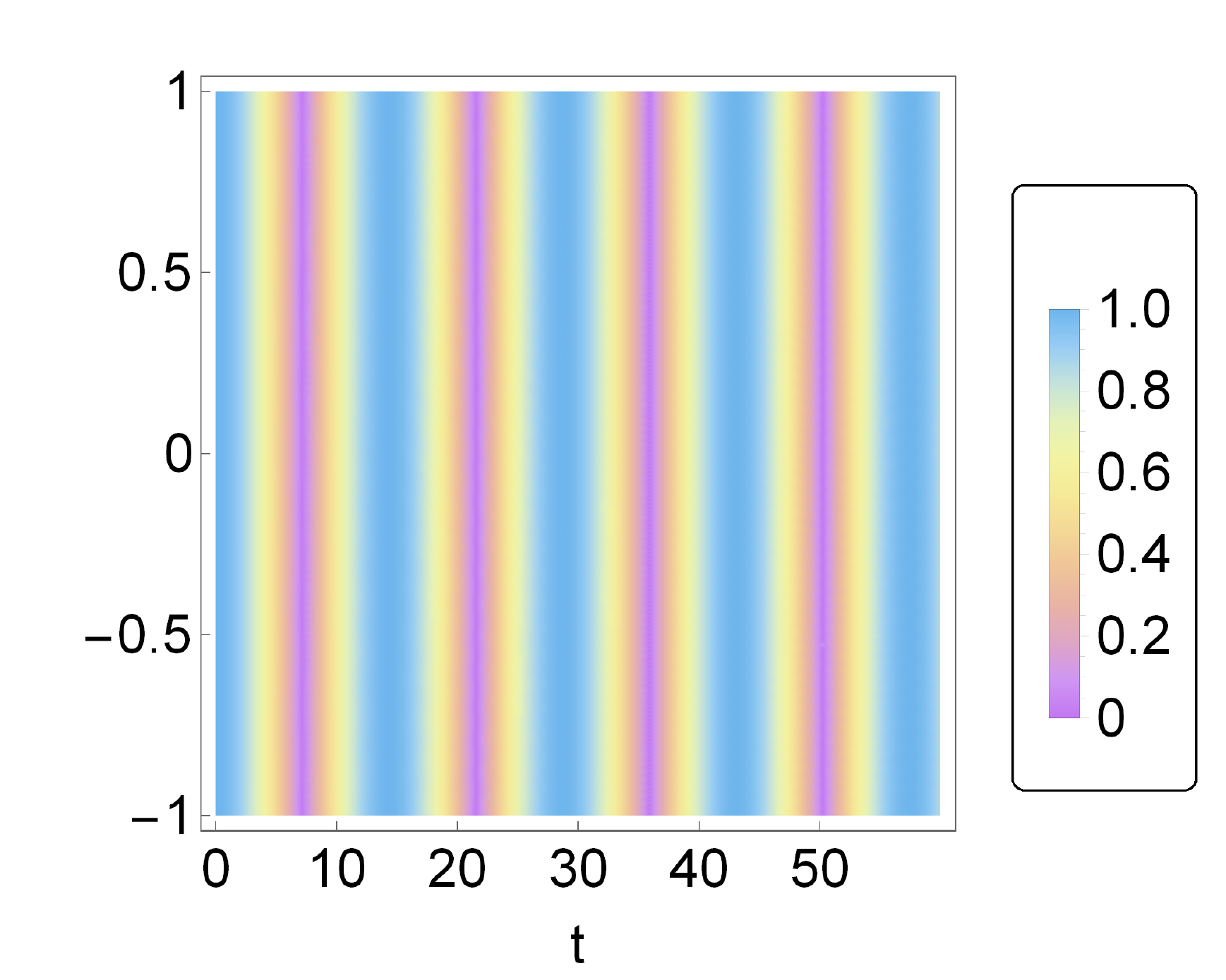}
\put(-155,60){\small $DM$}
\put(-90,110){$(c)$}
	\caption{The behavior of $\mathcal{S}_{or}(DM)$ for a system is initially prepared in a MES with ,g=0.1 $\lambda=1$, and $N=7$, where (a) $\gamma=0$, (b)$\gamma=0.5$ and (c)$\gamma=1$.}
	\label{g.9}
\end{figure}

Fig.(\ref{g.9}) shows the behavior of the orthogonality $\mathcal{S}_{or}(DM)$  as a contour plot, where we consider that the environment system consists of $7$ qubits. The small numbers of the orthogonality  are  displayed for the $XX$ chain model, namely  as one increases $\gamma$, the number of orthogonality decreases, and consequently the time of orthogonality increases. However, as we have discussed above, these results  will be changed if the initial environment's qubits are large. On the other hand, the largest tie of orthogonality is displayed for the Ising interaction model.

\begin{figure}
\centering
	\includegraphics[width=0.5\linewidth, height=6cm]{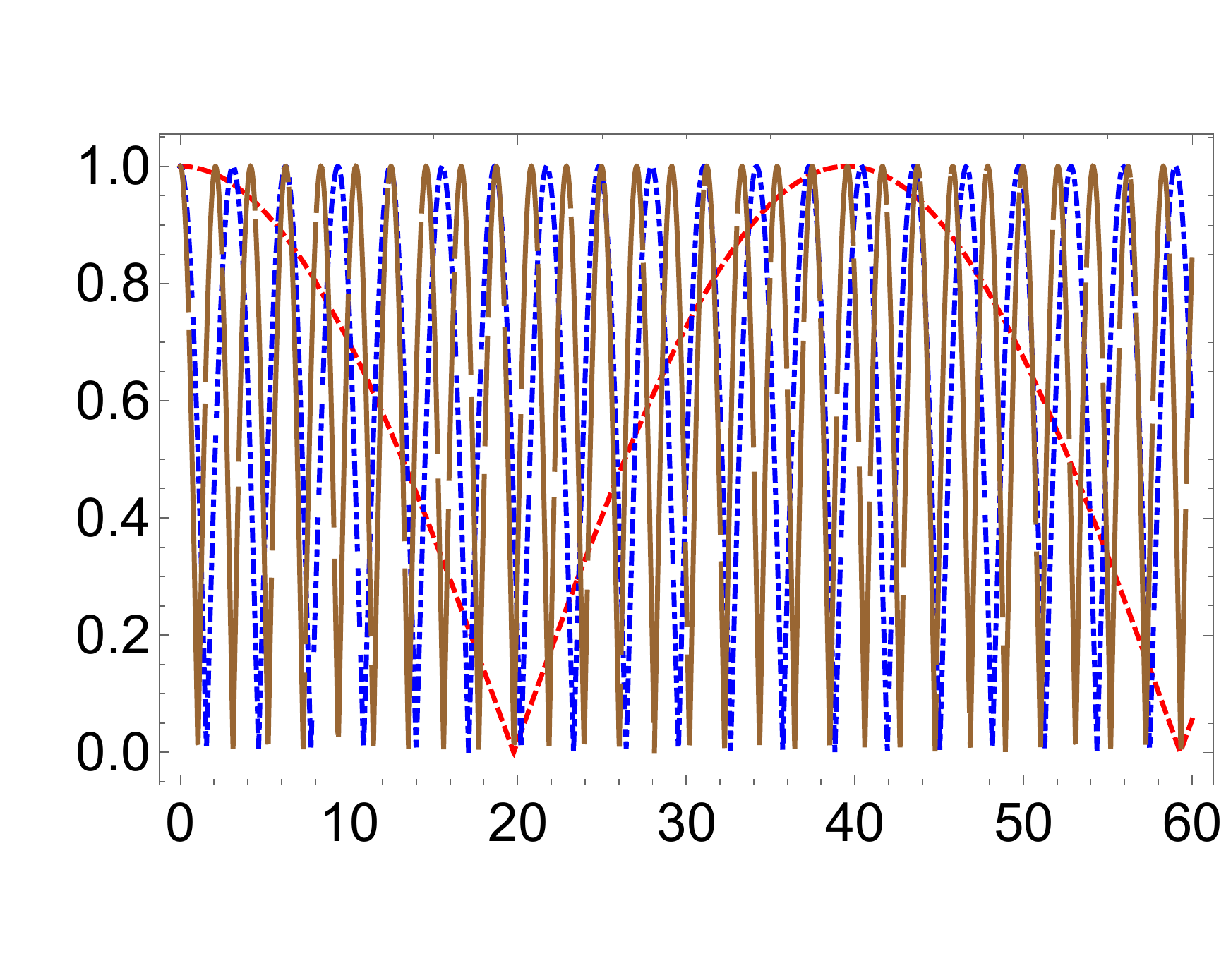}
\put(-240,90){$\mathcal{S}_{or}$}
\put(-110,10){$t$}
	\caption{The effect of the $DM$ interaction for MES   same as Fig.(2a) but we set $g=0.3$}
	\label{N.4}
\end{figure}

In Fig.(\ref{N.4}), we investigate the behavior of $\mathcal{S}_{or}(DM)$ for the different interaction types, where we assume that the number of the environment's qubits , $N=7$. It is clear that, the largest  number orthogonality are displayed at $\gamma=0$, namely the interaction is described by the $XX$ chain model. However the smallest numbers are predicted for the Ising model, where $\gamma=1$.  Moreover, the  shortest time of orthogonality and consequently the speed of transforming the information is shown for the $XX$ chain model.

\section{Conclusion}
In this manuscript, we discuss the orthogonality speed of a two qubit system interacting locally with different types of spin chain;  $XX$, Ising and the anistropic, in the presence of  Dzyaloshinsky-Moriya interaction. We assume that, the initial two qubit system is prepared in  a generic pure state.
 The analytical time evolution of the final state is obtained  as well as its components, namely its eigenvectors.
  The effect of the interaction's parameters and the initial state settings on the orthogonality speed is examined.

 Our results show that the numbers of environments' qubits play a central role on increasing the orthogonality numbers, where as one increases the site numbers  the orthogonality numbers increases and consequently the orthogonality time decreases. Moreover, the type of spin interaction can by used as a controller on the  orthogonality speed, where it is shown that the shortest orthogonality  time is displayed for the $XX$ spin chain, while the largest time is depicted when the Ising  interaction model is applied.
 The effect of the external field is examined in the presence of all the three different types of spin interaction. It is shown that in the absences of the external  field, the numbers of orthogonality  are small during  a period of interaction time. However, as one increases the strength of the external field the orthogonality's numbers  increase. Moreover, the difference between the orthogonality time that displayed for the three different spin interactions, may be minimized by increasing the strength  of the external field.
Furthermore, the behavior of the orthogonality is examined when the  interaction of Dzyaloshinsky-Moriya (DM) is switched on.  As it is shown from the listed figures, the numbers of orthogonality increase as one increases the strength of DM interaction.  However, this effect is clearly  displayed if  one increases the numbers of the sites' qubits of the  environment.

 Finally, the effect of the initial state settings on the orthogonality speed is investigated, where we consider that the initial system either prepared in a maximum, partial or separable state. The interaction's Hamilations parameters have the same effect for all the initial state settings. Moreover, for small values of the external field  strength, the maximum entangled state is robust against the decoherence  induced by the interaction of the qubit system with the qubit's environment. Therefore, the  orthogonality speed is larger than that displayed for the separable/partial entangled states.  However, as the  strength of the external field increases, the possibility  that the maximum entangled state losses its coherence increases.

{\it In conclusion:} the orthogonality time is examined for different initial state settings interacts  locally with different types of spin interaction models. It is shown that, the shortest time of orthogonality is displayed for the $XX$ chain model, while the largest time is shown for the Ising model. The external field increases the numbers of orthogonality, while Dzyaloshinsky-Moriya interaction decreases the  time of orthogonality. The initial state settings together with the external field have a significant effect on decreasing/increasing the time of orthogonality.
%

{}

\end{document}